\documentclass[runningheads]{llncs}
\usepackage{algorithm}
\usepackage{amsmath,amssymb,amsfonts,bbm}
\usepackage{algpseudocode}
\usepackage{graphicx}
\usepackage{textcomp}
\usepackage{blkarray}
\usepackage{longtable}
\usepackage{rotating}
\usepackage[dvipsnames]{xcolor}
\usepackage[english]{babel}
\usepackage{cite}

\usepackage{multirow,tabularx}
\newcolumntype{Y}{>{\centering\arraybackslash}X}

\newcommand{\coo}{\ensuremath{\mathrm{CO_2}}}

\begin{document}

\title{GResilience: Trading off between the Greenness and the Resilience of Collaborative AI Systems}
\titlerunning{GResilience: Trading off between the Greenness and the Resilience of CAIS \vspace{-15pt}}
%
\author{Diaeddin Rimawi\inst{1}\orcidID{0000-0003-3791-399X} \and
Antonio Liotta\inst{1}\orcidID{0000-0002-2773-4421} \and
Marco Todescato \inst{2}\orcidID{0000-0003-1449-5692} \and
Barbara Russo\inst{1}\orcidID{0000-0003-3737-9264}}
\authorrunning{D. Rimawi \textit{et al.}}
%
\institute{Free University of Bozen-Bolzano, Faculty of Engineering, Bolzano  39100, Italy 
\email{\{drimawi,antonio.liotta,barbara.russo\}@unibz.it}\\\and
Fraunhofer Italia, Bolzano 39100, Italy\\
\email{marco.todescato@fraunhofer.it}\vspace{-15pt}}
\maketitle              
\begin{abstract}
A Collaborative Artificial Intelligence System (CAIS) works with humans in a shared environment to achieve a common goal.  
To recover from a disruptive event that degrades its performance and ensures its resilience, a CAIS may then need to perform a set of actions either by the system, by the humans, or collaboratively together.
As for any other system, recovery actions may cause energy adverse effects due to the additional required energy. Therefore, it is of paramount importance to understand which of the above actions can better trade-off between resilience and greenness.
In this in-progress work, we propose an approach to automatically evaluate CAIS recovery actions for their ability to trade-off between the resilience and greenness of the system. We have also designed an experiment protocol and its application to a real CAIS demonstrator.
Our approach aims to attack the problem from two perspectives: as a one-agent decision problem through optimization, which takes the decision based on the score of resilience and greenness, and as a two-agent decision problem through game theory, which takes the decision based on the payoff computed for resilience and greenness as two players of a cooperative game.

 \vspace{-20pt}
\end{abstract}

\enlargethispage{\baselineskip}
\vspace{-10pt}
\section{Introduction}
\vspace{-5pt}
\label{sec:intro}

A Collaborative Artificial Intelligence System (CAIS) is an example of a Cyber Physical System that works together with humans in a shared environment to achieve a common goal, \cite{camilli_risk-driven_2021}. 
The collaboration between humans and  AI components poses specific challenges for a CAIS to be resilient (i.e., recover from a disruptive event that causes performance degradation) as disruptive events can be caused or have effects on humans. 
Therefore, it is of paramount importance to define suitable recovery strategies  to automatically support the decision-making process in case of disruptive events affecting CAISs. 
A recovery strategy typically detects the performance degradation (detection), then defines mitigation actions (mitigation), and finally, restores the system to an acceptable performance state (recovery), \cite{colabianchi_discussing_2021}.
Recovering from a disruptive event may require additional energy consumption, which, in turn, may increase the CAIS energy adverse effects, such as \coo{} footprint, \cite{kharchenko_concepts_2017, rimawi_green_2022}. 
The efficient usage of energy with minimizing adverse effects is called \textit{greenness}, \cite{kharchenko_concepts_2017}. 

In this in-progress work, we are interested in the relation between two properties of a CAIS: resilience and greenness. In particular, we aim to support the decision-making process to trade-off between them. 
Specifically, we introduce the approach \textit{GResilience}\footnote{The name GResilience comes from joining the two words, green and resilience. The name is inspired by the ``eco--greslient" technique used in \cite{mohammed_eco-gresilient_2018} study.} to select  automatically   recovery action(s) that finds the best trade-off between greenness and  resilience while restoring the CAIS services to an acceptable performance state. 
Our approach formulates the trade-off decision in two ways: i) as a one-agent decision through optimization, and ii) as a multi-agent decision through game theory. 
In the former case, the decision is taken by selecting actions(s) and optimizing measures of resilience and greenness. While in the latter, actions are selected through a multi-participant game in which the measures of resilience and greenness define the payoff for the actions.

Finally, we plan to apply our approach to a CAIS demonstrator available at our laboratory. The demonstrator is a robotic arm that is equipped with an AI component and performs activities for in-production systems.  
To this aim, we have devised an experimental protocol that we present in the next sections. 

In summary, we aim to \textit{understand the relationship between resilience and greenness and discuss the one-agent and the multi-agent methods in case of CAISs}. To achieve this goal, this study poses the following research questions:
\vspace{-5pt}
\begin{itemize}
    \item \textbf{RQ1}: Is the optimization model a valuable solution for automatizing the decision-making process that finds a trade-off between the \textit{greenness} and \textit{resilience} in CAISs?
    \item \textbf{RQ2}: Is the game theory model a valuable solution for automatizing the decision-making process that finds a trade-off between the \textit{greenness} and \textit{resilience} in CAISs?
    \item \textbf{RQ3}: What are the major differences between the optimization model and the game theory model solutions?
\end{itemize}

\vspace{-5pt}
The rest of this paper discusses the related work (Sec.~\ref{sec:relatedwork}), our approach and experiment protocol to the trade-off between resilience and greenness (GResilience) along with our demonstrator CAIS (Sec.~\ref{sec:approach}), and finally our conclusion and future work (Sec.~\ref{sec:conc}).
\enlargethispage{\baselineskip}
\vspace{-10pt}
\section{Related Work}
\vspace{-5pt}
\label{sec:relatedwork}
For what concerns our work, we see the following topics as relevant: i) resilience, ii) greenness, iii) multi-objective optimization, and iv) game theory.
 
\noindent\textbf{Resilience.} Studies are categorized into two classes depending on the model type they use, \cite{hosseini_review_2016}. The first category addresses quantitative models that discuss structure-based models and define computational metrics, whereas the second category addresses qualitative studies that are more concerned about conceptual frameworks. 
Henry \textit{et al.} \cite{henry_generic_2012} define a generic quantitative approach that uses a function of time to model the resilience process, while Speranza \textit{et al.} \cite{speranza_indicator_2014} propose a social-ecological framework to address policies' effectiveness to build livelihood resilience.

\noindent\textbf{Greenness.} Studies have analyzed greenness in two ways, technical and non-technical. Pandey \textit{et al.} \cite{pandey_greentpu_2020} discuss
 making Google Tensor Processing Unit resilient against activating sequences error in the systolic array considering a low-voltage operation, which ensures less energy adverse effects, while  Rodriguez \textit{et al.} \cite{rodriguez_green_2020} focuses on the business and financial aspect to identify the right location for a green infrastructure component of a sewer system. 

\noindent\textbf{Multi-Objective Optimization.} These techniques help create a simple mathematical representation of problems that have multiple objectives, \cite{gunantara_review_2018}. Several studies have used scalarization optimization techniques to trade-off between different system properties \cite{mohammed_eco-gresilient_2018}. 
Mohammed \textit{et al.} \cite{mohammed_eco-gresilient_2018} create the eco-gresilient model to build an economical, green, resilient supply chain network. It uses a three-objectives-optimization model (economical, green, resilient) to find the right number of facilities to be built in each supply network section. 
Multi-objective optimization helps find a solution that trades off multiple objectives and it uses one agent that combines the conflicting weighted objectives to find the final solution, \cite{gunantara_review_2018}. 

\noindent Finally, \textbf{Game Theory} searches for strategies (Nash equilibrium) that help the players gain the best payoffs for their interests, \cite{rimawi_green_2022, stowe_cheating_2010}. Thus, the problem is framed as a multi-agent game for which each agent has a preferred action to achieve a common goal  such as the game of ``The Battle of Sexes", \cite{stowe_cheating_2010}. The players of the game will try to choose the action that maximizes their payoff. When choosing this action is done independently  leads to a mixed strategy Nash equilibrium (MSNE), \cite{stowe_cheating_2010, rimawi_green_2022}.

\enlargethispage{\baselineskip}
\vspace{-10pt}
\section{Approach - GResilience}
\vspace{-5pt}
\label{sec:approach}
\textit{GResilience} is our empirical approach to support the decision-making process and trade-off between \textit{greenness} and \textit{resilience} after a disruptive event.
The approach provides one or more agents with the measurements required to take the decision and select the recovery action that best balance between the two properties.
The approach is applied to CAIS in which the AI component learns from human movements. GResilience monitors and controls the collaboration between the human and the AI component to support the decision-making process after disruptive events. 
The core common component of the GResilience approach is the measurement framework for resilience and greenness. The  framework is then used by two techniques to trade-off between greenness and resilience: optimization and game theory. While the former is typically used for trading off between non-functional properties of a system~\cite{mohammed_eco-gresilient_2018}, the latter, to the best of our knowledge, is novel in such context. 
The use of one or the other depends on the type of problem the decision-maker needs to solve.
The goal of each technique is to select a recovery action after the disruptive event to return to an acceptable performance state.
Recovery actions are categorized into two classes: i) general actions that are derived from a system or environment policies, and ii) actions defined by the decision maker. 

\begin{figure}
\includegraphics[width=\textwidth]{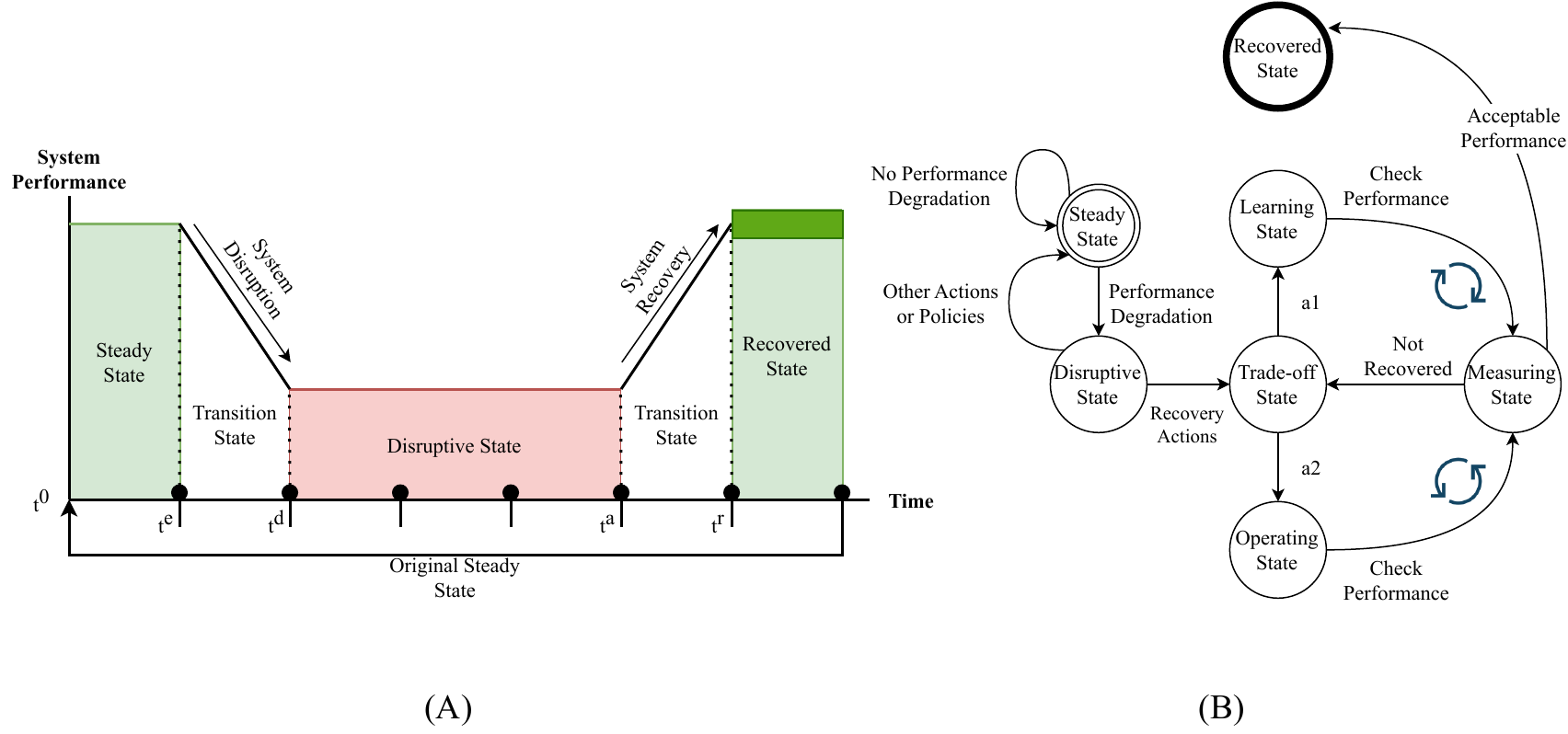}
\vspace{-20pt}
\caption{System's Performance States (A) Resilience performance evolution, (B) GResilience State Diagram} \label{fig:sysperstates}
\end{figure}

In Fig.~\ref{fig:sysperstates} we describe the resilience process from two perspectives. 
Fig.~\ref{fig:sysperstates}~(A) describes the performance behavior of the running system over time, while, Fig~\ref{fig:sysperstates}~(B), illustrates the GResilience approach state diagram based on such resilience process.
In Fig.~\ref{fig:sysperstates}~(A), the process starts at a \textit{steady state} and faces a disruptive event at $t^e$ that may \textit{transition} to a \textit{disruptive state} at $t^d$. During the disruptive state, the system starts the recovery process and selects a recovery action to move to an \textit{acceptable recovery state} at $t^r$.
In Fig~\ref{fig:sysperstates}~(B), the system starts at a \textit{steady state} and remains in the same state if there is no performance degradation. When a performance degradation occurs, the system moves to a \textit{disruptive state}, where it either recovers by default system actions or policies, or it moves to the \textit{trade-off state}. The trade-off state invokes the GResilience model (optimization or game theory) to select the action that finds the best trade-off between greenness and resilience. As an example in Fig.~\ref{fig:sysperstates}~(B), we illustrate two actions: a1) that targets a \textit{learning state}, and a2) that targets an\textit{operating state} of the AI component.
From either state, the performance recovery enters in the \textit{measuring state} looping between one state and the measurement state until the system reaches an acceptable performance (\textit{recovered state}).

\vspace{-20pt}
\vspace{-20pt}

The GResilience framework includes three attributes, one attribute to measure the system resilience and the other two to measure the CAIS greenness. The first attribute is \textit{the estimated run time} ($E_{t}$), which represents the action running time. The other two attributes are \textit{the estimated \coo{} footprint} ($E_{\coo{}}$) to emit by the action, and \textit{the human labor cost} ($H$), which is the number of human's interactions needed by the action. It is worth noticing here that human energy and financial factors are not considered in the framework and will be a matter of future investigation. The measures are then used by each of the two techniques as in the following.

\noindent\textbf{Optimization.} The optimization technique uses one agent and the \textit{weighted sum model} (WSM) \cite{mohammed_eco-gresilient_2018} to trade-off between greenness and resilience by combining the three attributes and define a global score for each action. Eq.~\eqref{eq:globalscore} shows the global score of the action $a$ ($S(a)$), where $w_{T}$, $w_{H}$, and $w_{\coo{}}$ are the weights of the attributes (run time, human labor, and \coo{} footprint).
$\epsilon$ is the confidence level of the AI component model ($\epsilon \in [0, 1]$): the higher the value the more we trust the AI to continue operating. Thus, $\epsilon$  multiplies the inverse of the resilience measure, and $1 - \epsilon$ multiplies the inverse of the greenness measures. Each resulting measure is then normalized (N()). Finally, we search the weights' values that maximize $S(a)$.
\vspace{-5pt}
\begin{equation}
\label{eq:globalscore}
        S(a) = w_{T} \cdot \epsilon \cdot N(E_{t}^{-1}) + (1 - \epsilon)\cdot\{ w_{H} \cdot N(H) + w_{\coo{}}  \cdot N(E_{\coo{}}^{-1})\}
        \vspace{-5pt}
\end{equation}

\noindent\textbf{Game Theory.} In the game theory technique, we leverage ``The Battle of Sexes" \cite{stowe_cheating_2010} and define our game \textit{The GResilience Game}.
The GResilience Game is played by two agents: $P_{g}$ and $P_{r}$. $P_{g}$ aims to make the system green by minimizing the \coo{} footprint by being more dependent on the human whereas $P_{r}$ aims to make the system resilient by minimizing the running time. Both players share the same goal of recovering the system so they need to adopt a strategy to recover the system and achieve both players' goals.
The  payoff matrix of the GResilience Game has  the same form  as in ``The Battle of Sexes",  Table~\ref{mat:payoffgeneral}. This table shows two Pure Strategies Nash Equilibria (PSNE) where both players choose the same action and a Mixed Strategy Nash Equilibrium (MSNE) based on the probability of each player's action, \cite{rimawi_green_2022, stowe_cheating_2010}.
Eq.~\eqref{eq:resiliencepayoff} shows the expressions to find the $P_{r}$ and $P_{g}$ payoffs,
where $\alpha$ is the matching factor that is $1$ in case the players land on different actions and $2$ in case of PSNE.

\begin{table}[t]
\caption{The GResilience Game General Payoff Matrix } \vspace{-20pt}
\centering
\scriptsize
\renewcommand{\arraystretch}{1.8}
\begin{longtable}[c]{l|l|c|c|c|}
     \multicolumn{2}{c}{} & \multicolumn{2}{c}{\pmb{$P_g$}} \\\cline{2-5}
    && \pmb{$a_1 (p)$} & \pmb{$a_2 (1-p)$} & \pmb{$P_r$  Expected Payoff} \\\cline{2-5}
     \multirow{2}{*}{\begin{sideways}\pmb{$P_r$}\end{sideways}}
     & \pmb{$a_1 (q)$} & $P^{2}_{r}(a_1), P^{2}_{g}(a_1)$ & $P^{1}_{r}(a_1), P^{1}_{g}(a_2)$ & $p P^{2}_{r}(a_1)  + (1 - p)  P^{1}_{r}(a_1)$\\\cline{2-5}
    & \pmb{$a_2 (1-q)$} & $P^{1}_{r}(a_2), P^{1}_{g}(a_1)$ & $P^{2}_{r}(a_2), P^{2}_{g}(a_2)$ & $p  P^{1}_{r}(a_2)  + (1 - p)  P^{2}_{r}(a_2)$ \\\cline{2-5}
    & \pmb{$P_g$  Expected Payoff} & $q  P^{2}_{g}(a_1)  + (1 - q)  P^{1}_{g}(a_1)$ & $q  P^{1}_{g}(a_2)  + (1 - q)  P^{2}_{g}(a_2)$ & \\\cline{2-5}
\end{longtable}
\label{mat:payoffgeneral}
\vspace{-20pt}
\end{table}

\vspace{-5pt}
\begin{equation}
\label{eq:resiliencepayoff}
   P^{\alpha}_{r}(a) = \epsilon \cdot \alpha \cdot E_{t}^{-1}, \,\,
   P^{\alpha}_{g}(a) = (1 - \epsilon) \cdot \alpha \cdot H^{-1} \cdot E_{\coo{}}^{-1}
\end{equation}
In the MSNE, $P_r$ chooses $a_1$ with probability $q$ and $a_2$ with probability $1-q$, while $P_g$ chooses $a_1$ with probability $p$ and $a_2$ with probability $1-p$, which results to the \textit{expected payoff} described in Table~\ref{mat:payoffgeneral}. Thus, to find the probability $q$ (resp. $p$) with MSNE, we equal the expected payoffs of $P_g$ (resp. $P_r$) for $a_1$ and $a_2$ and solve the resulting equation for $q$ (resp. $p$).

\noindent\textbf{Experiments Protocol.} We plan a series of experiments in three stages, i) setup, ii) iterative execution and data collection, and iii) data analysis. During the setup stage, we need to understand the disruptive events that might occur, and what are the feasible actions to recover from one of these events. As illustrated by Fig.~\ref{fig:sysperstates}~(B), GResilience wraps the system to detect the performance degradation in the second stage, and for each trading off technique, we collect the number of iterations to recover, the performance at the start and the end of each iteration, the selected action per iteration, and the values of the resilience and greenness attributes per iteration. Finally, by analyzing the collected data, we can understand for which disruptive event a technique is a valuable solution to automatize the decision-making process, and what are the major differences between them.

\begin{figure}
\includegraphics[width=\textwidth]{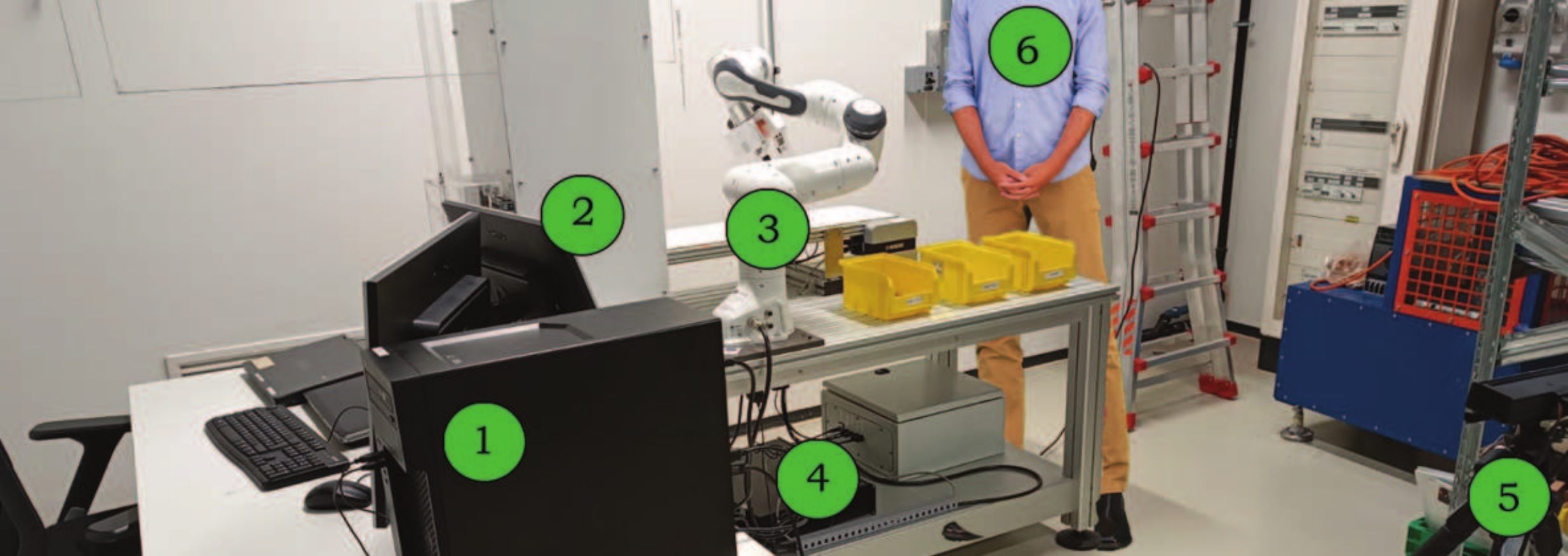}\centering
\caption{Collaborative Robot Learning from Demonstrations (CORAL)} 
\vspace{-20pt}
\label{fig:coral}
\end{figure}

\noindent\textbf{Demonstrator.} Our CAIS demonstrator ``CORAL"\footnote{CORAL is developed by Fraunhofer Italia Research in the context of ARENA Lab.} is a collaborative robot arm learning from demonstrations. 
Fig~\ref{fig:coral} shows the robotic arm, (where 1, 3, and 4 represent the arm and its controllers) that works with the human (6) to classify objects moving on the conveyor belt (2) based on their colors.
In addition to object color learning, CORAL learns background subtraction, object detection, and human movement. CORAL has two vision sensors, one through a Kinect (5) that monitors human movement, and a second above a conveyor belt that moves objects to be classified. 
Losing the lights that support the vision sensors or having another human in the vision range may disrupt CORAL ability to classify the objects and drop or wrongly classify them. Thus, CORAL requires more time to learn the objects with the faded environmental light and  this consumes additional energy. 
We will apply our approach to CORAL under different disruptive events. 

\vspace{-15pt}
\vspace{-10pt}

\enlargethispage{\baselineskip}
\section{Conclusion and Future Work}
\vspace{-10pt}
\label{sec:conc}
This in-progress work proposes an approach to support the decision-making process to recover from a disruptive event that has caused performance degradation and control the energy adverse effects at the same time. To this aim, we have defined a set of measures for resilience and greenness and two techniques leveraging optimization and game theory respectively.
The techniques automate the selection process of the recovery actions by measuring the trade-off between the greenness and the resilience capability of CAIS.
The first technique
evaluates each action separately using an optimization model (WSM), whereas the second technique evaluates greenness and resilience payoffs by selecting an action through a game theory model leveraging ``The Battle of Sexes".
To verify our approach, we designed experiments to test our techniques on our CAIS demonstrator.
In our future work, we plan to run experiments on CORAL. 
This will help us understand the relationship between resilience and greenness for CAIS and eventually extend our approach to test CAIS for other non-functional properties.
In addition, we plan to extend our techniques with reinforcement learning, to incorporate a rewarding mechanism in the optimization and game theory techniques. Moreover, we plan to reconsider further human attributes as for example human energy and financial costs.
\vspace{-10pt}

\bibliographystyle{splncs04}
\bibliography{references}

\end{document}